\documentclass[aps,prb,showpacs,twocolumn,floats]{revtex4}
\usepackage{amssymb}

\usepackage{graphicx,psfig}
\usepackage{dcolumn}
\usepackage{bm}

\input{tcilatex}

\begin{document}

\title{Phonon bottleneck in the low-excitation limit }
\author{D. A. Garanin}
\affiliation{\mbox{Department of Physics and Astronomy, Lehman
College, City University of New York,} \\ \mbox{250 Bedford Park
Boulevard West, Bronx, New York 10468-1589, U.S.A.} }

\date{\today}

\begin{abstract}
The phonon-bottleneck problem in the relaxation of two-level systems (spins)
via direct phonon processes is considered numerically in the weak-excitation
limit where the Schr\"{o}dinger equation for the spin-phonon system
simplifies. The solution for the relaxing spin excitation $p(t),$ emitted
phonons $n_{\mathbf{k}}(t),$ etc. is obtained in terms of the exact
many-body eigenstates. In the absence of phonon damping $\Gamma _{\mathrm{ph}%
}$ and inhomogeneous broadening, $p(t)$ approaches the bottleneck plateau $%
p_{\infty }>0$ with strongly damped oscillations, the frequency being
related to the spin-phonon splitting $\Delta $ at the avoided crossing. For
any $\Gamma _{\mathrm{ph}}>0$ one has $p(t)\rightarrow 0$ but in the case of
strong bottleneck the spin relaxation rate is much smaller than $\Gamma _{%
\mathrm{ph}}$ and $p(t)$ is nonexponential. Inhomogeneous broadening
exceeding $\Delta $ partially alleviates the bottleneck and removes
oscillations of $p(t).$ The line width of emitted phonons, as well as $%
\Delta ,$ increase with the strength of the bottleneck, i.e., with the
concentration of spins.
\end{abstract}
\maketitle


\section{Introduction}

\label{Sec-Intro}

Spin-lattice relaxation is an old an much studied problem that currently
recieves a resurge of attention because of its vital importance in quantum
information processing (see, e.g., Ref. \onlinecite{ardetal07prl} and
references therein). Theoretical description of the spin-lattice relaxation
as a single-spin process is in many cases insufficient because of the
collective effects of incoherent and coherent nature, such as the phonon
bottleneck \cite{vle41pr} and superradiance,\cite{dic54} respectively.

The problem of phonon bottleneck (PB) in relaxation of two-level systems
(henceforth spins) via direct phonon emission/absorption processes, first
recognized by Van Vleck\cite{vle41pr} in 1941, remains unsolved until now.
In two words, if the emitted phonons have nowhere to go, they are absorbed
by spins again and thus the spins cannot relax efficiently. However
transparent this picture might appear, is not easy to propose a theoretical
description of the effect based on the first principles.

Published theories of the PB \cite
{faustr61jpcs,scojef62pr,brywag67pr,abrble70,pinfai90prb} use \emph{ad hoc}
rate equations for populations of spins and resonant phonons, considering
the latter as a single dynamical variable. This is certainly an
oversimplification, because the emitted phonons, having frequencies $\omega
_{\mathbf{k}},$ form a group with a bell-like line shape with some width,
centered around the spin transition frequency $\omega _{0}.$ For a single
spin embedded into an infinite elastic matrix (as well as for a decaying
atomic state in a free space) this line shape is Lorentzian with the width $%
\Gamma /2,$ where $\Gamma $ is the single-spin decay rate following from the
Fermi golden rule. \cite{weiwig30a,hei54} However, in the case of many spins
with a concentration sufficient to create a bottleneck, the line shape and
line width of emitted phonons are unknown and should follow from the
solution of the problem.

Van Vleck came to the idea of the phonon bottleneck comparing the rate of
energy transfer from spins to phonons (obtained using experimental data)
with the phonon relaxation rate $\Gamma _{\mathrm{ph}}$ due to different
mechanisms and he found the latter to be typically too small to keep the
phonon subsystem at equilibrium. However, the primary role in the PB problem
belongs to another parameter that is not related to the phonon relaxation
rate. This parameter is of a statistical origin and is defined as the ratio
of the number of spins to the number of phonon modes within the single-spin
line width $\Gamma $. \cite{gar07prb} If this so-called bottleneck parameter
$B$ is vanishingly small, the spin excitation goes over into the phonon
subsystem and never returns. In this case spins completely relax even
without any phonon damping. However, for nonzero $B$ and $\Gamma _{\mathrm{ph%
}}=0$, the spin relaxation ends in the so-called bottleneck\ plateau that
corresponds to a quasiequilibrium between spins and resonant phonons but not
to the complete equilibrium. Further relaxation to the complete equilubrium
can be achieved only if $\Gamma _{\mathrm{ph}}$ is taken into account. It
should be stressed that the effective relaxation rate of the spins in this
case is not $\Gamma _{\mathrm{ph}}.$ It is much smaller and can be estimated
as $\Gamma _{\mathrm{ph}}$ multiplied by the small fraction of phonon modes
in the total number of modes (phonons + spins) involved in the process.

Although in many practical situations the number of resonant phonon modes is
determined by the inhomogeneous broadening of spin levels, the pure
spin-phonon model without inhomogeneous broadening has a fundamental
importance. It was shown \cite{gar07prb} that this model cannot be described
\emph{kinetically} (i.e., in terms of spin and phonon \emph{populations}
only) because of long-memory effects. In Ref. \onlinecite{gar07prb} memory
effects have been taken into account within a minimal approximation, adding
a new variable that can be interpreted as spin-phonon correlator. Analytical
and numerical solutions of the resulting \emph{reversible} \emph{dynamical}
equations show that the spin excitation approaches the botleneck plateau
with damped oscillations. Inclusion of an \emph{ad hoc} phonon damping $%
\Gamma _{\mathrm{ph}}$ into the bottleneck equations allows to describe the
second stage of the relaxation towards the complete equilibrium.

Still, the solution of the PB problem in Ref. \onlinecite{gar07prb} is not
completely satisfactory since it cannot produce a well-behaved line shape of
emitted phonons. This indicates that additional nondiagonal correlators
should be taken into account that will make the description more
complicated. Another important factor that should be taken into account is
the inhomogeneous broadening of the spin levels.

This paper presents the exact numerical solution of the phonon bottleneck
problem based on the Schr\"{o}dinger equation for the spin-phonon system,
with and without the \emph{ad hoc} phonon damping and inhomogeneous spin
broadening. The full Schr\"{o}dinger equation for a many-body system is, of
course, intractable by direct methods because of too many variables.
However, the \emph{low-excited} states of the system can be described by a
\emph{single excitation} that is hopping between spins and phonon modes. In
this case the Hilbert space of the problem is severely truncated and one has
to work with matrices the size of which is just the total number of spins
and phonon modes under consideration. This is the case considered here and
the solution of the PB problem is obtained by matrix algebra using Wolfram
Mathematica. The results of the calculations show that for the pure model
the spin excitation approaches the bottleneck plateau with oscillations,
however, less revealed than in Ref. \onlinecite{gar07prb}. The line width of
emitted phonons broadens with the bottleneck parameter $B.$

The structure of the rest of the paper is the following. Sec.\ \ref
{Sec-Hamiltonian} sets up the Hamiltonian of the spin-phonon system and the
Schr\"{o}dinger equation (SE) in the case of a single excitation. The spin
excitation and the initial conditions for problem are defined here. Sec.\ \ref
{Sec-NonB} presents the known results for the relaxation of a \emph{single
spin} and for the energy distribution of emitted phonons, used later for the
reference. Sec.\ \ref{Sec-Bottleneck} introduces the bottleneck parameter $B$
for systems of many spins from statistical arguments, both with and without
inhomogeneous spin broadening. Sec.\ \ref{Sec-DynamicMatrix} is the central
section of the paper introducing the matrix formalism for the
single-excitation spin-phonon problem. Here the expressions for the spin
excitation $p(t)$, its asymptotic value $p_{\infty }$ (the bottleneck
plateau), and the asymptotic populations of the emitted phonons are obtained
in terms of eigenvectors and eigenvalues of the dynamical matrix of the
system. General analysis of the eigenstates of the dynamical matrix is done
in Sec.\ \ref{Sec-Eigenstates}. It is shown here that the number of the
phonon modes ``on speaking terms'' with spins for the pure problem increases
with $B$ thus changing the statistical balance between spins and phonon
modes. In this section the formulas describing the spectrum of the \emph{%
split spin-phonon modes} and the hybridization of different phonon modes
with each other (i.e., the scattering of phonons on spins) are obtained in
terms of the eigenstates of the dynamical matrix. Sec.\ \ref{Sec-Numerical}
presents the results of numerical calculations for the pure model, including
the split spin-phonon modes with the gap, time evolution of the spin
excitation, bottleneck plateau, the post-plateau relaxation due to the
phonon damping, and the energy distribution of the emitted phonons in the
pure model. In Sec.\ \ref{Sec-Inhomo} the effects of the inhomogeneous spin
broadening are considered. The latter is shown to wash out oscillations of
the spin excitation $p(t).$ Sec.\ \ref{Sec-parameters} contains the
implementation of the general results to the spin relaxation between
adjacent spin levels in molecular magnets. In Sec.\ \ref{Sec-Discussion}
further problems of collective spin-phonons relaxation are discussed.

\section{The Hamiltonian and Schr\"{o}dinger equation}

\label{Sec-Hamiltonian}

Consider a spin-phonon Hamiltonian for $N_{S}$ two-level systems (spins) put
at positions $\mathbf{r}_{i}$ within an elastic body of $N$ cells
\begin{equation}
\hat{H}=\hat{H}_{0}+\hat{V},  \label{HamFull}
\end{equation}
where
\begin{equation}
\hat{H}_{0}=-\frac{1}{2}\sum_{i}\hbar \Bbb{\omega }_{0i}\sigma _{iz}+\sum_{%
\mathbf{k}}\hbar \omega _{\mathbf{k}}a_{\mathbf{k}}^{\dagger }a_{\mathbf{k}}
\label{H0Def}
\end{equation}
describes spins and harmonic phonons, $\mathbf{\sigma }$ being the Pauli
matrix. The spin transition frequencies $\Bbb{\omega }_{i}$ can differ from
site to site. One can represent them in the form
\begin{equation}
\Bbb{\omega }_{0i}=\overline{\omega _{0}}+\delta \omega _{0i},
\label{deltaomegaiDef}
\end{equation}
where $\delta \omega _{0i}\ll \overline{\omega _{0}}$ is the inhomogeneous
broadening. In the absense of the latter we will simply use $\omega _{0}$ as
the spin transition frequency. Neglecting the processes that do not conserve
the energy (that can be done in cases of practical significance where $\hat{V%
}$ can be treated as a perturbation), one can write $\hat{V}$ in the
rotating-wave approximation (RWA) as
\begin{equation}
\hat{V}=-\frac{\hbar }{\sqrt{N}}\sum_{i}\sum_{\mathbf{k}}\left( A_{i\mathbf{k%
}}^{\ast }X_{i}^{01}a_{\mathbf{k}}^{\dagger }+A_{i\mathbf{k}}X_{i}^{10}a_{%
\mathbf{k}}\right) ,  \label{VRWA}
\end{equation}
where $A_{i\mathbf{k}}\equiv V_{\mathbf{k}}e^{-i\mathbf{k\cdot r}_{i}}.$ In
the numerical work below $V_{\mathbf{k}}$ will be replaced by a constant, $%
V_{\mathbf{k}}\Rightarrow V$. The operator $X^{10}\equiv \sigma _{-}$ brings
the spin from the ground state $\left| \uparrow \right\rangle \equiv \left|
0\right\rangle \equiv \binom{1}{0}$ to the excited state $\left| \downarrow
\right\rangle \equiv \left| 1\right\rangle \equiv \binom{0}{1}$ while $%
X^{01}\equiv \sigma _{+}$ does the opposite. Note that the state with no
phonons and all spins in the ground state is the true ground state of the
Hamiltonian above.

Below we will consider the low-excited states of the spin-phonon system that
can be described by a superposition of the vacuum state of the system $%
\left| 0\right\rangle $ (no phonons and all spins in the ground state) and
the states with one excitation that is hopping between the spins and phonon
modes. The wave function of these states has the form
\begin{equation}
\Psi =\left( c_{0}+\sum_{i}c_{i}X_{i}^{10}+\sum_{\mathbf{k}}c_{\mathbf{k}}a_{%
\mathbf{k}}^{\dagger }\right) \left| 0\right\rangle ,  \label{PsiLowExc}
\end{equation}
where the coefficients satisfy the system of equations
\begin{equation}
\frac{dc_{0}}{dt}=i\omega _{0}c_{0}  \label{SEGSEq}
\end{equation}
and
\begin{eqnarray}
\frac{dc_{i}}{dt} &=&-i\delta \omega _{i}c_{i}+\frac{i}{\sqrt{N}}\sum_{%
\mathbf{k}}A_{i\mathbf{k}}c_{\mathbf{k}}  \nonumber \\
\frac{dc_{\mathbf{k}}}{dt} &=&-i\left( \omega _{\mathbf{k}}-\omega
_{0}\right) c_{\mathbf{k}}+\frac{i}{\sqrt{N}}\sum_{i}A_{i\mathbf{k}}^{\ast
}c_{i},  \label{SELowExc}
\end{eqnarray}
up to an irrelevant global phase factor. One can see that the ground-state
coefficient $c_{0}$ is decoupled from the other coefficients since the RWA
Hamiltonian conserves the excitation number
\begin{equation}
p(t)+\sum_{\mathbf{k}}n_{\mathbf{k}}(t)=\mathrm{const},  \label{ExcCons}
\end{equation}
where
\begin{equation}
p=\sum_{i}\left| c_{i}\right| ^{2},\qquad n_{\mathbf{k}}=\left| c_{\mathbf{k}%
}\right| ^{2}  \label{pDef}
\end{equation}
are the excitation number of the spin subsystem and populations of the
phonon modes. One can also define transverse spin polarization components by
\begin{eqnarray}
\left\langle \sigma _{+}\right\rangle &=&\left\langle X^{01}\right\rangle
=c_{0}^{\ast }\sum_{i}c_{i}.  \nonumber \\
\left\langle \sigma _{-}\right\rangle &=&\left\langle \sigma
_{+}\right\rangle ^{\ast },\qquad \left\langle \sigma _{x}\right\rangle =%
\func{Re}\left\langle \sigma _{+}\right\rangle ,
\label{TransverseComponentsDef}
\end{eqnarray}
etc. The main part of the time dependence of $\left\langle \sigma
_{+}\right\rangle $ is $e^{-i\omega _{0}t}.$ The absolute value of the
transverse spin component
\begin{equation}
\left\langle \sigma _{\bot }\right\rangle \equiv \sqrt{\left\langle \sigma
_{x}\right\rangle ^{2}+\left\langle \sigma _{y}\right\rangle ^{2}}=\left|
\left\langle \sigma _{+}\right\rangle \right|  \label{sigmaperpDef}
\end{equation}
does not have this oscillating factor. In many practical situations $%
\left\langle \sigma _{\bot }\right\rangle $ decays with time due to the
inhomogeneous broadening. In the absence of the latter, the only source of
the decoherence is interaction with phonons.

Our task is to find the time evolution $p(t)$ and $\left\langle \sigma
_{\bot }\right\rangle _{t}$ starting from a particular initial state. In
this work we restrict ourselves to the initial states with no phonons, $c_{%
\mathbf{k}}(0)=0$. The simplest initial condition in this case is one spin
at site $i_{0}$ excited and all other spins in their ground states:
\begin{equation}
c_{i_{0}}(0)=1,\qquad c_{i\neq i_{0}}(0)=0.  \label{ICOneSpinExc}
\end{equation}
Another kind of the initial spin state is the state with the excitation
equidistributed over all spins:
\begin{equation}
c_{i}(0)=e^{i\phi _{i}}/\sqrt{N_{S}}.  \label{ICEquidistr}
\end{equation}
The initial spin state with random phases
\begin{equation}
\left\langle e^{i(\phi _{i}-\phi _{j})}\right\rangle =\delta _{ij},
\label{RandomPhases}
\end{equation}
is called incoherent. If $\phi _{i}$ are constant or they periodically
change in space with some wave vector $\mathbf{q}_{0},$ the initial state is
coherent. One can consider other kinds of spin initial conditions, say,
excitation distributed over spins in some compact region of space.

\section{Non-bottlenecked spin-lattice relaxation}

\label{Sec-NonB}

The results of this section can be found in the literature,\cite
{weiwig30a,hei54} still a concise description of the non-bottlenecked
spin-phonon dynamics is presented for the sake of consistency and future
reference.

\subsection{Relaxation of a single spin}

\label{Sec-SingleSpin}

Suppose there is a single spin, $c_{i}=c,$ in the initially excited state.
With a proper choice of the origin of the coordinate system one has $A_{i%
\mathbf{k}}=V_{\mathbf{k}}.$ Using the Schr\"{o}dinger equation (\ref
{SELowExc}), one can integrate the equations for the phonon modes $c_{%
\mathbf{k}}$:
\begin{eqnarray}
c_{\mathbf{k}}(t) &=&\frac{iV_{\mathbf{k}}^{\ast }}{\sqrt{N}}%
\int_{t_{0}}^{t}dt^{\prime }e^{-i\left( \omega _{\mathbf{k}}-\omega
_{0}\right) (t-t^{\prime })}c(t^{\prime })  \nonumber \\
&=&\frac{iV_{\mathbf{k}}^{\ast }}{\sqrt{N}}\int_{0}^{t-t_{0}}d\tau
e^{-i\left( \omega _{\mathbf{k}}-\omega _{0}\right) \tau }c(t-\tau )
\label{ckIntegrated}
\end{eqnarray}
and insert the result into the equation for the spin $c$:
\begin{equation}
\frac{dc}{dt}=-\frac{1}{N}\sum_{\mathbf{k}}\left| V_{\mathbf{k}}\right|
^{2}\int_{0}^{t-t_{0}}d\tau e^{-i\left( \omega _{\mathbf{k}}-\omega
_{0}\right) \tau }c(t-\tau ).
\end{equation}
In this integro-differential equation, $c(t-\tau )$ is a slow function of
time, whereas the memory function $f(\tau )=\left( 1/N\right) \sum_{\mathbf{k%
}}\left| V_{\mathbf{k}}\right| ^{2}e^{-i\left( \omega _{\mathbf{k}}-\omega
_{0}\right) \tau }$ is sharply peaked at $\tau =0.$ Thus one can replace $%
c(t-\tau )\Rightarrow c(t),$ after which integration over $\tau $ and
keeping only real contribution responsible for the relaxation yields the
equation
\begin{equation}
\frac{dc}{dt}=-\frac{\Gamma }{2}c,  \label{cEqRelax}
\end{equation}
where
\begin{equation}
\Gamma =\frac{2\pi }{N}\sum_{\mathbf{k}}\left| V_{\mathbf{k}}\right|
^{2}\delta \left( \omega _{\mathbf{q}}-\omega _{0}\right)
\label{GammaSingSp}
\end{equation}
is the single-spin decay rate. For $V_{\mathbf{k}}=V$ independently of the
direction of $\mathbf{k,}$ $\Gamma $ can be written as
\begin{equation}
\Gamma =2\pi \left| V\right| ^{2}\rho _{\mathrm{ph}}\left( \omega
_{0}\right) ,  \label{Gammaviarho}
\end{equation}
where
\begin{equation}
\rho _{\mathrm{ph}}\left( \omega \right) =\frac{1}{N}\sum_{\mathbf{k}}\delta
\left( \omega _{\mathbf{k}}-\omega \right)  \label{rhoomegaDef}
\end{equation}
is the phonon density of states normalized by one. The accuracy of the above
short-memory approximation is justified by $\Gamma \ll \omega _{0}.$ The $%
\delta $-function in Eq.\ (\ref{GammaSingSp}) implies that the spin is
relaxing to a large number of phonon modes so that summation is replaced by
integration,
\begin{equation}
\frac{1}{N}\sum_{\mathbf{k}}\ldots \Longrightarrow \int \frac{d^{d}k}{\left(
2\pi \right) ^{d}}\ldots ,  \label{Sum2Int}
\end{equation}
where $d$ is spatial dimension. In small bodies with essentially discrete
phonon modes Eq.\ (\ref{Sum2Int}) is invalid. The solution of Eq.\ (\ref
{cEqRelax}) with $t_{0}=0$ is $c(t)=e^{-\left( \Gamma /2\right) t}$ that
leads to well-known decay rule for the spin excitation
\begin{equation}
p(t)=e^{-\Gamma t}.  \label{GoldenRelaxation}
\end{equation}
The rate of transverse spin relaxation according to Eqs.\ (\ref
{TransverseComponentsDef}) and (\ref{cEqRelax}) is $\Gamma /2.$

A similar elimination of phonons can be performed in the case of many spins
in the absence of the PB. \cite{chugar04prl} The resulting equations
describe collective spin-phonon relaxation, including superradiance. \cite
{dic54}

\subsection{Distribution of emitted phonons}

\label{Sec-EmittedPhonons}

After the time dependence $c(t)$ has been found, one can return to Eq.\ (\ref
{ckIntegrated}) and calculate $c_{\mathbf{q}}(t).$ With $t_{0}=0$ the result
is
\begin{equation}
c_{\mathbf{k}}(t)=\frac{iV_{\mathbf{k}}^{\ast }}{\sqrt{N}}\frac{e^{-i\left(
\omega _{\mathbf{k}}-\omega _{0}\right) t}-e^{-\Gamma t/2}}{-i\left( \omega
_{\mathbf{k}}-\omega _{0}\right) +\Gamma /2}.
\end{equation}
This leads to the distribution of emitted phonons
\begin{equation}
n_{\mathbf{k}}(t)=\left| c_{\mathbf{k}}(t)\right| ^{2}=\frac{\left| V_{%
\mathbf{k}}\right| ^{2}}{N}\frac{1-2e^{-\Gamma t/2}\mathrm{cos}\left[ \left(
\omega _{\mathbf{k}}-\omega _{0}\right) t\right] +e^{-\Gamma t}}{\left(
\omega _{\mathbf{k}}-\omega _{0}\right) ^{2}+\Gamma ^{2}/4}
\label{nktonespin}
\end{equation}
that asymptotically becomes the Lorentzian function
\begin{equation}
n_{\mathbf{k}}=\frac{1}{N}\frac{\left| V_{\mathbf{k}}\right| ^{2}}{\left(
\omega _{\mathbf{k}}-\omega _{0}\right) ^{2}+\Gamma ^{2}/4}.
\label{nkLorentz}
\end{equation}
With the help of Eqs.\ (\ref{GoldenRelaxation}) and (\ref{nktonespin}) one
can check that the total excitation is conserved, in accordance with Eq.\ (%
\ref{ExcCons}). For For $V_{\mathbf{k}}=V$ Eq.\ (\ref{nkLorentz}) can be
rewritten as
\begin{equation}
n_{\mathbf{k}}=\frac{1}{\pi N\rho _{\mathrm{ph}}(\omega _{0})}\frac{\Gamma /2%
}{\left( \omega _{\mathbf{k}}-\omega _{0}\right) ^{2}+\Gamma ^{2}/4}.
\label{nkLorentzGamma}
\end{equation}

\section{The phonon bottleneck}

\label{Sec-Bottleneck}

Let us now turn to systems with a macroscopic number of spins $N_{S}.$ At
least in the case of diluted spins, the relaxation is controlled by the
bottleneck parameter $B$ that can be defined as the ratio of the number of
spins $N_{S}$ to the number of phonon modes $N_{\Gamma }$ within the natural
spin line width $\Gamma $ of Eq.\ (\ref{GammaSingSp})$,$
\begin{equation}
N_{\Gamma }=\pi N\rho _{\mathrm{ph}}\left( \omega _{0}\right) \Gamma ,
\label{NGammaDef}
\end{equation}
where $\rho _{\mathrm{ph}}\left( \omega \right) $ is given by Eq.\ (\ref
{rhoomegaDef}). That is,\cite{gar07prb}
\begin{equation}
B\equiv \frac{N_{S}}{N_{\Gamma }}=\frac{N_{S}}{\pi N\rho _{\mathrm{ph}%
}\left( \omega _{0}\right) \Gamma }=\frac{n_{S}}{\pi \rho _{\mathrm{ph}%
}\left( \omega _{0}\right) \Gamma },  \label{BDef}
\end{equation}
$n_{S}$ $=N_{S}/N$ being the number of spins per unit cell. The definitions
above pertain to a single phonon branch and extension to several phonon
branches is obvious. For $B\lesssim 1$ (see below) $N_{\Gamma }$ is the
estimation of the number of phonon modes that can exchange excitation with
spins.

In the case of a single spin, $N_{S}=1,$ in a macroscopic ($N\rightarrow
\infty $) matrix the parameter $B$ is vanishingly small. The excitation,
initially localized at the spin, spreads with time over a large number $%
N_{\Gamma }$ of resonant phonon modes, so that the spin relaxes completely
according to Eq.\ (\ref{GoldenRelaxation}). In simulations, the macroscopic
limit is achieved if the average distance between the neighboring phonon
modes becomes smaller than the natural line width $\Gamma $
\begin{equation}
\frac{1}{N\rho \left( \omega _{0}\right) }\ll \Gamma .
\label{MacroscopicPhonons}
\end{equation}
If the sample is so small that the spin can exchange excitation with only a
few phonon modes, $N_{\Gamma }\sim 1,$ it does not relax completely. In this
case one has $B\sim 1,$ the so-called phonon bottleneck situation. The
simplest realization of the bottleneck is a system of two resonant states in
which the excitation oscillates between the two states in time. If $N_{S}$
is macroscopic but still $N_{S}\ll N_{\Gamma }$ and thus $B\ll 1,$ the
initial spin excitation is transferred irreversibly into the phonon
subsystem, as is clear from statistical arguments. There is no bottleneck
for $B\ll 1,$ and the spin relaxation is still described by Eq.\ (\ref
{GoldenRelaxation}).

In the case of a finite concentration of spins $n_{S},$ the parameter $B$
can easily become large. In this case only a small fraction of the
excitation migrates into the phonon subsystem and the spins practically
cannot relax as the emitted phonons are being absorbed by spins again. For $%
B\gg 1,$ as we will immediately see, the number of the phonon modes ``on
speaking terms'' with spins is not $N_{\Gamma }$ but much greater. The
latter can be obtained if one considers the spin-phonon hybridization.
Inserting the spin Fourier components $b_{\mathbf{k}}\equiv \sum_{i}c_{i}e^{i%
\mathbf{k\cdot r}_{i}}$ into the Schr\"{o}dinger equation (\ref{SELowExc})
and neglecting the coupling of modes with different values of $\mathbf{k,}$
i.e., using $\sum_{i}\nu _{i}e^{i\left( \mathbf{k-q}\right) \mathbf{\cdot r}%
_{i}}\Rightarrow N_{S}\delta \left( \mathbf{k-q}\right) ,$ one can reduce
Eq.\ (\ref{SELowExc}) to a 2$\times 2$ matrix problem that can be easily
diagonalized. The eigenstates of this problem are hybridized spin-phonon
modes with frequencies\cite{jacste63pr,calchugar07prb}
\begin{equation}
\Omega _{\mathbf{k}}^{(\pm )}=\frac{1}{2}\left\{ \omega _{\mathbf{k}}-\omega
_{0}\pm \sqrt{\left( \omega _{\mathbf{k}}-\omega _{0}\right) ^{2}+\Delta _{%
\mathbf{k}}^{2}}\right\} ,  \label{OmegapmDef}
\end{equation}
where
\begin{equation}
\Delta _{\mathbf{k}}=2\sqrt{n_{S}}\left| V_{\mathbf{k}}\right|
\label{Deltansk}
\end{equation}
is the spin-phonon splitting. These modes have the form of a straight line
with a slope corresponding to phonons and a horizontal line corresponding to
spins. The lines have an avoided level crossing at $\omega _{\mathbf{k}%
}=\omega _{0},$ split by $\Delta _{\mathbf{k}}$ that depends on the spin
concentration. For $V_{\mathbf{k}}=V$ independently of the direction of $%
\mathbf{k,}$ one can eliminate $\rho _{\mathrm{ph}}\left( \omega _{0}\right)
$ from Eq.\ (\ref{BDef}) with the help of Eqs.\ (\ref{rhoomegaDef}) and (\ref
{GammaSingSp}) and obtain the relation
\begin{equation}
B=\frac{2n_{S}\left| V\right| ^{2}}{\Gamma ^{2}}=\frac{\Delta ^{2}}{2\Gamma
^{2}}.  \label{BviaV2}
\end{equation}

For $\Delta \gtrsim \Gamma $ the number of phonon modes strongly coupled to
spins can be estimated as
\begin{equation}
N_{\Delta }=\pi N\rho _{\mathrm{ph}}\left( \omega _{0}\right) \frac{\Delta }{%
2}.  \label{NDeltaDef}
\end{equation}
If $n_{S}$ is so small that $\Delta $ falls below the natural spin line
width $\Gamma ,$ one cannot speak of the hybridyzed spin-phonon modes. From
Eq.\ (\ref{BviaV2}) one obtains
\begin{equation}
N_{\Delta }=N_{\Gamma }\sqrt{B/2}.  \label{NDeltaviaNGamma}
\end{equation}
The number of resonant phonons $N_{\mathrm{res}}$ that exchange excitation
with the spins can be estimated in the whole range of $B$ as
\begin{equation}
N_{\mathrm{res}}=\left\{
\begin{array}{cc}
N_{\Gamma }, & B\lesssim 1 \\
N_{\Delta }, & B\gtrsim 1.
\end{array}
\right.  \label{NresDef}
\end{equation}

For any $B>0$, the spin excitation $p(t)$ does not relax to zero but reaches
a plateau at some $p_{\infty }$ that from the statistical equidistribution
argument can be estimated as
\begin{equation}
p_{\infty }=\frac{N_{S}}{N_{S}+N_{\mathrm{res}}}  \label{pinfEstimation}
\end{equation}
with $N_{\mathrm{res}}$ defined above. We will see that this formula works
well both for $B\lesssim 1$ and $B\gtrsim 1,$ in the absence of the
inhomogeneous broadening that will be considered in Sec.\ \ref{Sec-Inhomo}.
The asymptotes of the above expression are
\begin{equation}
p_{\infty }\cong \left\{
\begin{array}{ll}
B, & B\ll 1 \\
1-1/\sqrt{2B}, & B\gg 1,
\end{array}
\right.  \label{pinfBlarge}
\end{equation}
in accordance with the numerical results of Sec.\ \ref{Sec-Numerical}.

To describe the complete spin relaxation after the bottleneck plateau, one
has to include the phonon relaxation processes, the easiest way being
ascribing an empirical relaxation rate $\Gamma _{\mathrm{ph}}$ to the
phonons.

\section{Dynamic matrix and the time evolution of the system}

\label{Sec-DynamicMatrix}

For the numerical solution of Eq.\ (\ref{SELowExc}) it is convenient to
introduce the state vector $\mathbf{C}=\left( \{c_{i}\},\{c_{\mathbf{k}%
}\}\right) $ and rewrite Eq.\ (\ref{SELowExc}) in the form
\begin{equation}
\frac{d\mathbf{C}}{dt}=-i\mathbf{\Phi \cdot C,}  \label{SELowExcMatr}
\end{equation}
where $\mathbf{\Phi }$ is the dynamical matrix of the spin-phonon system. In
the absence of phonon damping, $\mathbf{\Phi }$ is Hermitean. Since the
number of discreet phonon modes is $N,$ $\mathbf{\Phi }$ is a $\left(
N_{S}+N\right) \times \left( N_{S}+N\right) $ matrix. There are three
methods of numerical solution of this equation that can be implemented in
the Wolfram Mathematica.

The first method is the direct numerical solution using one of the
ordinary-differential equations (ODE) solvers. This method is fast, can be
made much faster if Mathematica is replaced by one of programming languages,
it does not require high accuracy, but it does not allow to analytically
average over the random phases in Eq.\ (\ref{ICEquidistr}). This averaging
can only be done if one runs the calculation many times with different
realizations of initial conditions.

The second method is based upon numerical calculation of the matrix
exponentials in the solution of Eq.\ (\ref{SELowExcMatr}) $\mathbf{C}(t)=e^{-i%
\mathbf{\Phi }t}\mathbf{C}(0)\mathbf{.}$ This method is slower than the
direct ODE solution, it also does not require high accuracy, and here one
can average over the initial conditions analytically.

The third method uses the expansion of the solution $\mathbf{C}(t)$ over
eigenvectors of $\mathbf{\Phi .}$ This method allows analytical averaging
over initial conditions, it is faster than the method using matrix
exponentials, if formulated in a fully vectorized form. However, this method
requires high precision and for large matrices it runs on 64-bit machines
only. Note that arbitrary-precision computations on 32-bit machines with
Mathematica are possible but they are very slow. An important advantage of
this method is that it allows to obtain formulas for the asymptotic $%
t\rightarrow \infty $ state of the system in terms of matrices.

Below, the method based on the eigenvectors of $\mathbf{\Phi }$ will be
used. In the general case when the \emph{ad hoc} phonon damping $\Gamma _{%
\mathrm{ph}}$ is added, $\mathbf{\Phi }$ is non-Hermitean, and one has to
distinguish between right and left eigenvectors. The dynamical matrix $%
\mathbf{\Phi }$ has $N_{S}+N$ right eigenvectors $\mathbf{R}_{\mu }$ that
satisfy
\begin{equation}
\mathbf{\Phi \cdot R}_{\mu }=\left( \Omega _{\mu }-i\Gamma _{\mu }\right)
\mathbf{R}_{\mu }.  \label{EigenFDef}
\end{equation}
In the eigenvalues, the imaginary parts $\Gamma _{\mu }$ originate from $%
\Gamma _{\mathrm{ph}}$. The size of the matrix $\mathbf{\Phi }$ can be
reduced if the phonon modes far from the resonance are dropped. The solution
of Eq.\ (\ref{SELowExcMatr}) can be expanded over the complete orthonormal
set of $\mathbf{R}_{\mu }$ as follows
\begin{equation}
\mathbf{C}(t)=\sum_{\mu }\mathbf{R}_{\mu }e^{-\left( i\Omega _{\mu }+\Gamma
_{\mu }\right) t}\mathbf{L}_{\mu }\cdot \mathbf{C}(0),  \label{CtSol}
\end{equation}
where $\mathbf{L}_{\mu }$ are left eigenvectors of $\mathbf{\Phi }$ that
satisfy $\mathbf{L}_{\mu }\cdot \mathbf{R}_{\nu }=\delta _{\mu \nu }.$ Note
that, in general, $\mathbf{R}_{\mu }$ and $\mathbf{L}_{\mu }$ are not
complex conjugate. The vectorized form of Eq.\ (\ref{CtSol}) is
\begin{equation}
\mathbf{C}(t)=\mathbf{E}\cdot \mathbf{W}(t)\cdot \mathbf{E}^{-1}\cdot
\mathbf{C}(0),  \label{CtSolVectorized}
\end{equation}
where $\mathbf{E}$ is the right-eigenvector matrix composed of all
eigenvectors $\mathbf{R}_{\mu }$ standing vertically, $\mathbf{E}^{-1}$ is
left-eigenvector matrix, composed of all left eigenvectors lying
horizontally, and $\mathbf{W}(t)$ is the diagonal matrix with the elements $%
e^{-\left( i\Omega _{\mu }+\Gamma _{\mu }\right) t}$. In fact, $\mathbf{E}%
\cdot \mathbf{W}(t)\cdot \mathbf{E}^{-1}=e^{-i\mathbf{\Phi }t}.$

\subsection{Longitudinal relaxation of spins}

\label{Sec-RelSpin}

Now the spin excitation $p(t)$ can be written with the help of Eqs.\ (\ref
{pDef}) and (\ref{CtSolVectorized}) in the vectorized form
\begin{equation}
p(t)=\left( \mathbf{E}\cdot \mathbf{W}(t)\cdot \mathbf{E}^{-1}\cdot \mathbf{C%
}(0)\right) _{S}\cdot \left( \mathrm{h.c.}\right) _{S},  \label{ptVectorized}
\end{equation}
where the subscript $S$ means projection onto the spin subspace. Eq.\ (\ref
{ptVectorized}) can be used for a fast computation. For the incoherent
initial condition, Eqs.\ (\ref{ICEquidistr}) and (\ref{RandomPhases}), one
obtains
\begin{equation}
p(t)=\frac{1}{N_{S}}\mathrm{Tr}_{S}\left[ \left( \mathbf{E}\cdot \mathbf{W}%
(t)\cdot \mathbf{E}^{-1}\right) _{S}\cdot \left( \mathrm{h.c.}\right) _{S}%
\right] ,  \label{ptVectorizedRandomPhase}
\end{equation}
where the trace is taken over spin indices only. A more explicit form of Eq.\
(\ref{ptVectorized}) is
\begin{eqnarray}
p(t) &=&\sum_{\mu \nu }e^{-\left( \Gamma _{\mu }+\Gamma _{\nu }\right)
t}\cos \left[ \left( \Omega _{\mu }-\Omega _{\nu }\right) t\right]
\sum_{n=1}^{N_{S}}R_{\mu n}^{\ast }R_{\nu n}  \nonumber \\
&&\times \sum_{n^{\prime }=1}^{N_{S}}L_{\nu n^{\prime }}C_{n^{\prime
}}(0)\sum_{n^{\prime \prime }=1}^{N_{S}}L_{\mu n^{\prime \prime }}^{\ast
}C_{n^{\prime \prime }}^{\ast }(0).  \label{ptNonVectorized}
\end{eqnarray}
In the absence of phonon damping, $\Gamma _{\mu }=0,$ there is a nonzero
asymptotic value of $p$ that can be obtained from the equation above by
dropping all oscillating terms, i.e., setting $\mu =\nu .$ This corresponds
to the diagonal density matrix of the spin-phonon system. Taking into
account $L_{\mu n}^{\ast }=R_{\mu n}$ in the Hermitean case, one obtains
\begin{equation}
p_{\infty }=\sum_{\mu }\sum_{n=1}^{N_{S}}\left| R_{\mu n}\right| ^{2}\left|
\sum_{n^{\prime }=1}^{N_{S}}R_{\mu n^{\prime }}^{\ast }C_{n^{\prime
}}(0)\right| ^{2}.  \label{pinfGeneral}
\end{equation}
For the incoherent initial condition this simplifies to
\begin{equation}
p_{\infty }=\frac{1}{N_{S}}\sum_{\mu }\left( \sum_{n=1}^{N_{S}}\left| R_{\mu
n}\right| ^{2}\right) ^{2}.  \label{pinfRandomPhase}
\end{equation}
Note that in the macroscopic limit $N\rightarrow \infty $ for a single spin
Eq.\ (\ref{ptNonVectorized}) should assume the simple exponential form of Eq.\
(\ref{GoldenRelaxation}) and $p_{\infty }=0$. The same should be the case
for any finite $N_{S}.$

If the phonon damping $\Gamma _{\mathrm{ph}}$ is finite but small, the
process of spin relaxation is two-stage. First, the spin subsystem
equilibrates with the subsystem of resonant phonons and $p(t)$ is mainly
changing due to the time dependence of the terms with $\mu \neq \nu $ in Eq.\
(\ref{ptNonVectorized}), whereas the role of $\Gamma _{\mu }$ is
insignificant. At the end of this stage the terms with $\mu \neq \nu $ die
out, and the further slow relaxation is governed by $\Gamma _{\mu }.$ In
particular, for the incoherent initial condition Eq.\ (\ref{ptNonVectorized})
at the second stage of the relaxation becomes
\begin{equation}
p(t)=\frac{1}{N_{S}}\sum_{\mu }e^{-2\Gamma _{\mu }t}\sum_{n=1}^{N_{S}}\left|
R_{\mu n}\right| ^{2}\sum_{n^{\prime }=1}^{N_{S}}\left| L_{\mu n^{\prime
}}\right| ^{2}.  \label{ptSpreadRandom2DampedSecond}
\end{equation}
Since for macroscopic systems the number of different values of $\Gamma
_{\mu }$ in this expression is very large, the dependence $p(t)$ is a
combination of many different exponentials, i. e., $p(t)$ is nonexponential.

In the limit we study in the paper, $k_{0}r_{0}\gg 1,$ the solution $p(t)$
is actually the same for coherent and incoherent initial conditions. The
only difference is that for the incoherent initial condition averaging over
the initial phases of spins leads to reproducible results for different
realizations of the nondiagional elements of $\mathbf{\Phi .}$ For coherent
initial conditions, one obtains somewhat different results for different
realizations of the spin-phonon matrix elements, related to location of the
individual spins in space. These differences persist with increasing the
number of spins and phonon modes, so that there is no self-avereaging. The
computation of $p(t)$ in the coherent case is faster but averaging over spin
configurations is needed.

\subsection{Transverse relaxation of spins}

\label{Sec-RelSpin-Trans}

Let us now consider the time evolution of the transverse spin polarization
given by Eq.\ (\ref{sigmaperpDef}) starting from the fully coherent initial
condition
\begin{equation}
c_{i}=\frac{\sin \theta }{\sqrt{N_{S}}},\qquad c_{0}=\cos \theta
\label{TransIniCond}
\end{equation}
that satisfies the normalization condition $\left| c_{0}\right|
^{2}+\sum_{i}\left| c_{i}\right| ^{2}=1$ for the wave function of Eq.\ (\ref
{PsiLowExc}) in the case of initial phonon vacuum. In Eq.\ (\ref{TransIniCond}%
) $\theta $ is the angle between the spin vector and the $z$-axis. In the
initial state one has
\begin{equation}
\left\langle \sigma _{\bot }\right\rangle _{0}=\sqrt{N_{S}}\sin \theta \cos
\theta  \label{sigmaTr0}
\end{equation}
that can be used to define the normalized transverse spin polarization
\begin{equation}
f_{\bot }(t)=\left\langle \sigma _{\bot }\right\rangle _{t}/\left\langle
\sigma _{\bot }\right\rangle _{0}  \label{fTrDef}
\end{equation}
that satisfies $f_{\bot }(0)=1.$ With the help of Eq.\ (\ref{CtSolVectorized}%
) one obtains the vectorized expression
\begin{equation}
f_{\bot }(t)=\left| \frac{1}{N_{S}}\sum_{nn^{\prime }=1}^{N_{S}}\left(
\mathbf{E}\cdot \mathbf{W}(t)\cdot \mathbf{E}^{-1}\right) _{nn^{\prime
}}\right| .  \label{fTrRes}
\end{equation}
Note that this formula is explicitly independent of the angle $\theta .$ An
alternative expression for $f_{\bot }(t)$ following from Eq.\ (\ref{CtSol})
has the form
\begin{equation}
f_{\bot }(t)=\left| \sum_{\mu }T_{\mu }e^{-\left( i\Omega _{\mu }+\Gamma
_{\mu }\right) t}\right| ,  \label{fTrResAlt}
\end{equation}
where
\begin{equation}
T_{\mu }\equiv \frac{1}{N_{S}}\sum_{n=1}^{N_{S}}R_{\mu n}\sum_{n^{\prime
}=1}^{N_{S}}L_{\mu n^{\prime }}.  \label{TmuDef}
\end{equation}

In the case of undamped phonons one has $L_{\mu n}^{\ast }=R_{\mu n}$ and $%
T_{\mu }=\left( 1/N_{S}\right) \left| \sum_{n=1}^{N_{S}}R_{\mu n}\right|
^{2} $ is real. Then, dropping oscillating terms in Eq.\ (\ref{fTrResAlt}) at
large times one obtains the asymptotic value $f_{\bot }(\infty )=\sqrt{%
\sum_{\mu }\left| T_{\mu }\right| ^{2}}$ [c.f. transition from Eq.\ (\ref
{ptNonVectorized}) to Eq.\ (\ref{pinfGeneral})]. It can be easily shown that
in the presence of inhomogeneous broadening and in the absence of the
coupling to phonons $f_{\bot }(\infty )\rightarrow 0$ in the thermodynamic
limit. The same should hold in the presence of both inhomogeneous broadening
and spin-phonon interaction since the former should be sufficient to cause
complete decoherence. Analysis of the principally important case without the
inhomogeneous broadening should be postponed until obtaining numerical
results. For a single spin, coupling to phonons causes decoherence with the
rate $\Gamma /2,$ as was stressed at the end of Sec.\ (\ref{Sec-SingleSpin}).
In the case of the phonon bottleneck this decoherence should be slowed down
since a few phonon modes couple to many spins, still the expected result is $%
f_{\bot }(\infty )=0.$

\subsection{Energy distribution of emitted phonons}

\label{Sec-EmittedPhononsNum}

The method formulated above can be used to find the state of the phonon
system resulting from spin-phonon relaxation. Obviously it can be done for
undamped phonons only. In the case of many spins there are no analytical
results for the occupation numbers of emitted phonons $n_{\mathbf{k}}.$ in
the literature. However, one can express $n_{\mathbf{k}}$ through the
eigenstates of the dynamical matrix $\mathbf{\Phi }$ defined by Eqs.\ (\ref
{SELowExcMatr}) and (\ref{EigenFDef}). For the wave function given by Eq.\ (%
\ref{PsiLowExc}) one has $n_{\mathbf{k}}=\left| c_{\mathbf{k}}\right| ^{2}.$
Labeling the phonon modes by the discrete index $l$, one can express $n_{%
\mathbf{k}}\equiv n_{l}$ through the state vector $\mathbf{C}$ given by Eq.\ (%
\ref{CtSol}). For the incoherent initial spin state with the help of Eq.\ (%
\ref{RandomPhases}) one obtains, asymptotically,
\begin{equation}
n_{l}(\infty )=\frac{1}{N_{S}}\sum_{\mu }\sum_{n=1}^{N_{S}}\left| R_{\mu
n}\right| ^{2}\left| R_{\mu ,N_{S}+l}\right| ^{2}.  \label{nkinfSpreadRandom}
\end{equation}
For $N_{S}=1$ this expression should reproduce Eq.\ (\ref{nkLorentz}).

\section{Analysis of the eigenstates of the spin-phonon system}

\label{Sec-Eigenstates}

\subsection{Spinness and off-resonance phonon emission}

Although the formalism of the preceding section is sufficient to describe
the dynamics of the spin-phonon system in terms of transition between
different bare (unperturbed) modes, it is also interesting to look closer at
the true spin-phonon eigenstates. Throughout this section we consider
phonons as undamped, $L_{\mu n}^{\ast }=R_{\mu n}$. The eigenstates are
superpositions of spin and phonon states, so the first question would be to
determine the fractions of spin and phonon states in any eigenstate, or, as
it can be termed, their ``spinness'' and ``phononness''. For instance, the
spinness of the state $\mu $ is defined by
\begin{equation}
Spinness_{\mu }=\sum_{n=1}^{N_{S}}\left| R_{\mu n}\right| ^{2},
\label{SpinnessDef}
\end{equation}
while the phononness is defined by a similar expression with summation over
phonon indices. Note that spinness enters the expression for the asymptotic
spin excitation $p_{\infty },$ Eq.\ (\ref{pinfRandomPhase}). Obviously the
sum of spinness and phononness of any state $\mu $ is 1. Spinness summed
over $\mu $ gives the total number of spins. Far from the resonance the
eigenstates are mainly phonon states, so that their spinness is small and
can be calculated perturbatively. Labeling these states by the wave vector $%
\mathbf{k}$ instead of $\mu ,$ one has $R_{\mathbf{qk}}\cong \delta _{%
\mathbf{qk}}$, where $\delta $ is the Kronecker symbol, and, in the first
order of the perturbation theory,
\begin{equation}
R_{\mathbf{k}i}\cong -\frac{1}{\sqrt{N}}\frac{A_{i\mathbf{k}}}{\omega _{%
\mathbf{k}}-\omega _{0}}.  \label{RkiPert}
\end{equation}
With this one obtains
\begin{equation}
Spinness_{\mathbf{k}}\cong \frac{n_{S}\left| V_{\mathbf{k}}\right| ^{2}}{%
\left( \omega _{\mathbf{k}}-\omega _{0}\right) ^{2}}\Longrightarrow \frac{B}{%
2}\frac{\Gamma ^{2}}{\left( \omega _{\mathbf{k}}-\omega _{0}\right) ^{2}}.
\label{SpinnessNonRes}
\end{equation}
The last expression was obtained for $\left| V_{\mathbf{k}}\right|
^{2}=\left| V\right| ^{2}$ with the help of Eq.\ (\ref{BviaV2}). One can see
that for a large bottleneck parameter the interaction of phonons with spins
becomes large, so that phonon modes are noticeably distorted even relatively
far from the resonance. This means, dynamically, that the number of phonon
modes that exchange excitation with spins is not just a fixed number $%
N_{\Gamma }$ determined by the single-spin relaxation rate $\Gamma $ [see
Eq.\ (\ref{NGammaDef})] and it increases with $B.$ Of course, Eq.\ (\ref
{SpinnessNonRes}) becomes inapplicable near the resonance. However, one can
figure out its behavior for $B\gg 1.$ In this case the spinness of the
states near the resonance should approach 1. One can reproduce this behavior
by adding $\left( B/2\right) \Gamma ^{2}$ in the denominator of Eq.\ (\ref
{SpinnessNonRes}). This is in accordance with Eqs.\ (\ref{NDeltaviaNGamma})
and (\ref{NresDef}) that define the number of phonon modes on speakng terms
with the spins.

For $B\ll 1$ we will see that the spinness remains small everywhere
including the resonance. This means that initially prepared states with the
excitation localized on spins will decompose over the true eigenstates that
have a very small fraction of spin states, so that the excitation will
migrate completely into the phonon subsystem. To the contrary, for $B\gg 1$
the initially prepared state will decompose over eigenstates that have
spinness close to 1, so that the excitation mostly remains on spins that is
the phonon bottleneck.

Using Eqs.\ (\ref{RkiPert}) and (\ref{nkinfSpreadRandom}) allows one to
obtain the probability of emission of a phonon with a wave vector $\mathbf{k}
$ far from the resonance. With $\mu \Rightarrow \mathbf{q}$ one obtains
\begin{eqnarray}
n_{\mathbf{k}} &=&\frac{1}{N_{S}}\sum_{\mathbf{q}}\sum_{i=1}^{N_{S}}\left|
R_{\mathbf{q}i}\right| ^{2}\left| R_{\mathbf{qk}}\right| ^{2}  \nonumber \\
&\cong &\frac{1}{N_{S}}\sum_{i=1}^{N_{S}}\left| R_{\mathbf{k}i}\right| ^{2}=%
\frac{1}{N}\frac{\left| V_{\mathbf{k}}\right| ^{2}}{\left( \omega _{\mathbf{k%
}}-\omega _{0}\right) ^{2}},  \label{nkoffres}
\end{eqnarray}
independently of the bottleneck parameter $B$. This result is in accord with
Eq.\ (\ref{nkLorentz}).

\subsection{Spin-phonon hybridization}

To quantify the spin-phonon hybridization in the general case, one can
introduce the average phonon detuning from the spins $\omega _{\mathbf{k}%
}-\omega _{0}$ in any eigenstate $\mu $ as
\begin{equation}
\left\langle \omega _{\mathbf{k}}-\omega _{0}\right\rangle _{\mu }\equiv
\frac{\sum_{\mathbf{k}}\left| R_{\mu \mathbf{k}}\right| ^{2}\left( \omega _{%
\mathbf{k}}-\omega _{0}\right) }{\sum_{\mathbf{k}}\left| R_{\mu \mathbf{k}%
}\right| ^{2}},  \label{kmuDef}
\end{equation}
where the denominator is the phononness of the eigenstate $\mu $ introduced
above. Note that this definition does not include the momentum carried by
spins. Thus Eq.\ (\ref{kmuDef}) principally cannot completely reproduce the
results for dense magnetic systems. Plotting $\left\langle \omega _{\mathbf{k%
}}-\omega _{0}\right\rangle _{\mu }$ vs the energy (frequency) eigenvalues $%
\Omega _{\mu }$ reveals that near the resonance there are pairs of the same
detunings $\left\langle \omega _{\mathbf{k}}-\omega _{0}\right\rangle _{\mu
} $ for two different frequencies $\Omega _{\mu }$. In other words, the
frequencies corresponding to the same $\left\langle \omega _{\mathbf{k}%
}-\omega _{0}\right\rangle _{\mu }$ are split because of the spin-phonon
interaction. Far from the resonance the relation between $\left\langle
\omega _{\mathbf{k}}-\omega _{0}\right\rangle _{\mu }$ and the frequency $%
\Omega _{\mu }$ corresponds to a pure phonon mode.

\subsection{Resonance scattering of phonons}

One also can study the admixture of other phonon modes to the given phonon
mode because of the phonon scattering on spins. This can be described by the
dispersion
\begin{equation}
\delta \left\{ \omega _{\mathbf{k}}-\omega _{0}\right\} _{\mu }\equiv \sqrt{%
\frac{\sum_{\mathbf{k}}\left| R_{\mu \mathbf{k}}\right| ^{2}\left( \omega _{%
\mathbf{k}}-\omega _{0}-\left\langle \omega _{\mathbf{k}}-\omega
_{0}\right\rangle _{\mu }\right) ^{2}}{\sum_{\mathbf{k}}\left| R_{\mu
\mathbf{k}}\right| ^{2}}}.  \label{deltahmuDef}
\end{equation}
Far from the resonance the eigenstates are almost pure phonons, so that $%
\delta \left\{ \omega _{\mathbf{k}}-\omega _{0}\right\} _{\mu }$ is small.
At resonance $\delta \left\{ \omega _{\mathbf{k}}-\omega _{0}\right\} _{\mu
} $ has a maximum that corresponds to the resonance scattering.

\section{Numerical results and analysis}

\label{Sec-Numerical}

This section has been omitted to reduce the size. Please, get the full text of the paper here:\\

www.lehman.edu/faculty/dgaranin/Bottleneck2.pdf

\section{Inhomogeneous broadening of spin levels}

\label{Sec-Inhomo}

In some systems the inhomogeneous broadening defined by Eq.\ (\ref
{deltaomegaiDef}) is much larger than the natural spin linewidth $\Gamma $
or the spin-phonon splitting $\Delta .$ The number of spins within the
frequency interval $d\omega _{0}$ around the frequency $\omega _{0}$ is
\begin{equation}
dN_{S}=N_{S}\rho _{S}(\omega _{0})d\omega _{0},  \label{rhoSDef}
\end{equation}
where the spin density of states $\rho _{S}(\omega _{0})$ satisfies $%
\int_{0}^{\infty }\rho _{S}(\omega _{0})d\omega _{0}=1.$ An example is the
Gaussian line shape
\begin{equation}
\rho _{S}(\omega _{0})=\frac{1}{\sqrt{2\pi }\delta \omega _{0}}\exp \left[ -%
\frac{\left( \omega _{0}-\overline{\omega _{0}}\right) ^{2}}{2\left( \delta
\omega _{0}\right) ^{2}}\right] .  \label{rhoSGaussian}
\end{equation}
In the presence of a large inhomogeneous broadening the number of phonons on
speaking terms with spins can be estimated as $N_{\delta \omega _{0}}\sim
N\pi \rho _{\mathrm{ph}}(\omega _{0})\delta \omega _{0}$ that should replace
$N_{\Gamma }$ of Eq.\ (\ref{NGammaDef}) in the definition of the bottleneck
parameter $B$ that becomes smaller than for the pure model by a factor of $%
\delta \omega _{0}/\Gamma \gg 1.$ Consideration of this kind can be found in
Van Vleck's original paper\cite{vle41pr} and subsequent publications. In
fact, as we shall see below in this section, the relation between the
numbers of spins and phonon modes that can exchange excitation is different
in different frequency regions within the inhomogeneous spin line width $%
\delta \omega _{0}$ and a frequency-resolved description of the phonon
bottleneck is possible.

\begin{figure}[t]
\includegraphics[angle=-90,width=8cm]{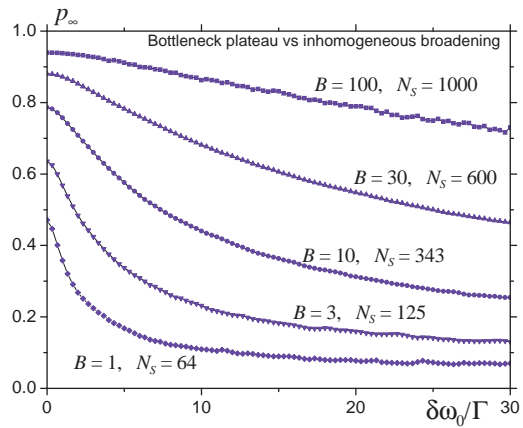}
\caption{The bottleneck plateau in the case of inhomogeneous broadening with
the Gaussian line shape vs $\protect\delta \protect\omega _{0}$.}
\label{Fig-Bottleneck_plateau-inhomo}
\end{figure}

In simulations the macroscopic limit is achieved if the average distance
between the neigboring spin levels becomes smaller than the natural line
width $\Gamma $
\begin{equation}
\frac{1}{N_{S}\rho _{S}(\omega _{0})}\ll \Gamma ,  \label{MacroscopicSpins}
\end{equation}
c.f. Eq.\ (\ref{MacroscopicPhonons}). Since both of these conditions should
be satisfied, both $N$ and $N_{S}$ should be large enough. The results for
the bottleneck plateau $p_{\infty }$ vs the inhomogeneous broadening $\delta
\omega _{0}$ in Eq.\ (\ref{rhoSGaussian}), obtained numerically from Eq.\ (\ref
{pinfRandomPhase}) are shown in Fig.\ \ref{Fig-Bottleneck_plateau-inhomo}.
Since the number of phonons that can exchange energy with spins increases
with $\delta \omega _{0},$ the bottleneck plateau $p_{\infty }$ decreases.
For $B\lesssim 1,$ the most pronounced decrease occurs at $\delta \omega
_{0}\sim \Gamma ,$ the crossover from the natural line width to the
inhomogeneous line width. In the case $B\gg 1$ this crossover occurs at $%
\delta \omega _{0}\sim \sqrt{B}\Gamma \sim \Delta ,$ in accordance with the
comments after Eq.\ (\ref{SpinnessNonRes}).

Clearly for $\delta \omega _{0}\gg \Gamma ,\Delta $ much more phonons can
exchange their energy with spins, so that the bottleneck condition is
alleviated and spin relaxation is facilitated. In this case the problem
simplifies since spins and phonons exchange energy only within the frequency
interval of order $\max (\Gamma ,\Delta )$ that is much narrower than the
inhomogeneous line width $\delta \omega _{0}$. Thus one can split the latter
into the frequency intervals $\Delta \omega _{0}$ around $\omega _{0}$ that
satisfy $\max (\Gamma ,\Delta )\ll $ $\Delta \omega _{0}\ll \delta \omega
_{0}$ and consider the energy exchange between spins and phonons in each of
these intervals independently. The spin and phonon densities of states
within each interval $\Delta \omega _{0}$ can be considered as constants and
they define the bottleneck parameter in a frequency interval around $\omega
_{0}$
\begin{equation}
B_{\omega _{0}}\equiv \frac{N_{S}}{N}\frac{\rho _{S}(\omega _{0})}{\rho _{%
\mathrm{ph}}\left( \omega _{0}\right) }=n_{S}\frac{\rho _{S}(\omega _{0})}{%
\rho _{\mathrm{ph}}\left( \omega _{0}\right) }.  \label{Bomega0Def}
\end{equation}
Since the phonon density of states is a smooth function, one can replace $%
\rho \left( \omega _{0}\right) \Rightarrow \rho \left( \overline{\omega _{0}}%
\right) .$ On the other hand, $\rho _{S}(\omega _{0})$ and thus $B_{\omega
_{0}}$ have a maximum at $\omega _{0}=\overline{\omega _{0}}.$ One can
parametrize
\begin{equation}
B_{\omega _{0}}=\frac{\rho _{S}(\omega _{0})}{\rho _{S}(\overline{\omega _{0}%
})}B_{\overline{\omega _{0}}},  \label{BomegaoviaBhatomega}
\end{equation}
where $B_{\overline{\omega _{0}}}$ is the bottleneck parameter at the center
of the line. For a Gaussian line shape of Eq.\ (\ref{rhoSGaussian}) one has
\begin{equation}
B_{\overline{\omega _{0}}}=\frac{n_{S}}{\sqrt{2\pi }\rho _{\mathrm{ph}%
}\left( \omega _{0}\right) \delta \omega _{0}}.  \label{BomegaAvrGauss}
\end{equation}
The spin excitation reaches a frequency-dependent plateau $p_{\infty
}(\omega _{0})$ that depends on $B_{\omega _{0}}.$ The average over the spin
line shape $\rho _{S}(\omega _{0})$ of Eq.\ (\ref{rhoSDef}) now becomes
\begin{equation}
\overline{p_{\infty }}=\int_{0}^{\infty }d\omega _{0}\rho _{S}(\omega
_{0})p_{\infty }(\omega _{0}).  \label{pinfAvDef}
\end{equation}
Evidently $\overline{p_{\infty }}<p_{\infty }(\overline{\omega _{0}})$ \
since the bottleneck effect weakens away from the center of the spin band.
The numerically found dependence of $p_{\infty }(\omega _{0})$ is very close
to
\begin{equation}
p_{\infty }(\omega _{0})=\frac{B_{\omega _{0}}}{1+B_{\omega _{0}}}
\label{pinfAvrRes}
\end{equation}
that can be expected from general statistical arguments. This important
formula will be used below to obtain results for the bottleneck plateau
taking into account the inhomogeneously broadened spin line shape. The
illustrations will be done for the Gaussian line shape of Eq.\ (\ref
{rhoSGaussian}).

\begin{figure}[t]
\includegraphics[angle=-90,width=8cm]{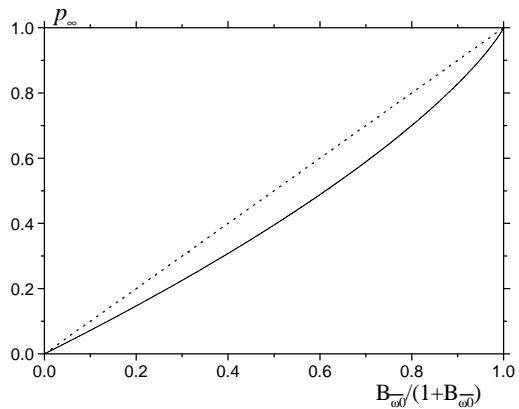}
\caption{The bottleneck plateau in the case of strong ($\protect\delta
\protect\omega _{0}\gg \Gamma, \Delta $) inhomogeneous broadening with the Gaussian
line shape vs the bottleneck parameter $B_{\protect\omega _{0}}$.}
\label{Fig-pinfInhomo}
\end{figure}

In the case $B_{\overline{\omega _{0}}}\ll 1$ one obtains
\begin{equation}
\overline{p_{\infty }}\cong B_{\overline{\omega _{0}}}\int_{0}^{\infty
}d\omega _{0}\frac{\left[ \rho _{S}(\omega _{0})\right] ^{2}}{\rho _{S}(%
\overline{\omega _{0}})}
\end{equation}
that with the help of Eq.\ (\ref{rhoSGaussian}) yields
\begin{equation}
\overline{p_{\infty }}\cong B_{\overline{\omega _{0}}}/\sqrt{2}.
\label{pinfomegaBomegasmall}
\end{equation}
If $B_{\overline{\omega _{0}}}\gg 1,$ one can use
\begin{equation}
\overline{p_{\infty }}=1-\int_{0}^{\infty }d\omega _{0}\frac{\rho
_{S}(\omega _{0})}{1+B_{\omega _{0}}}  \label{pinfomega0largeBomega}
\end{equation}
that follows from Eq.\ (\ref{pinfAvrRes}). The integrand of this expression
is close to $\rho _{S}(\overline{\omega _{0}})/B_{\overline{\omega _{0}}}=%
\mathrm{const}$ for $B_{\omega _{0}}\gtrsim 1$ and it decays abruptly
further from the center of the spin line where $B_{\omega _{0}}\lesssim 1.$
Let us define $\omega ^{\ast }$ that satisfies $B_{\omega ^{\ast }}=1,$ that
is, $\rho _{S}(\omega ^{\ast })=\rho _{S}(\overline{\omega _{0}})/B_{%
\overline{\omega _{0}}}.$ Then from Eq.\ (\ref{pinfomega0largeBomega}) one
obtains, approximately,
\begin{equation}
\overline{p_{\infty }}\cong 1-\frac{2\delta \omega ^{\ast }\rho _{S}(%
\overline{\omega _{0}})}{B_{\overline{\omega _{0}}}},
\end{equation}
where $\delta \omega ^{\ast }\equiv \left| \omega ^{\ast }-\overline{\omega
_{0}}\right| .$ For the Gaussian line shape of Eq.\ (\ref{rhoSGaussian}) one
has $\delta \omega ^{\ast }=\delta \omega _{0}\sqrt{2\ln B_{\overline{\omega
_{0}}}}$ and, finally,
\begin{equation}
\overline{p_{\infty }}\cong 1-\frac{2}{B_{\overline{\omega _{0}}}}\sqrt{%
\frac{\ln B_{\overline{\omega _{0}}}}{\pi }},\qquad B_{\overline{\omega _{0}}%
}\gg 1.  \label{pinfInhomoLargeB}
\end{equation}
Fig.\ \ref{Fig-pinfInhomo} shows $\overline{p_{\infty }}$ in the whole
interval of $B_{\overline{\omega _{0}}}$ calculated numerically from Eqs.\ (%
\ref{rhoSGaussian}), (\ref{pinfAvDef}), and (\ref{BomegaoviaBhatomega}).

\begin{figure}[t]
\includegraphics[angle=-90,width=8cm]{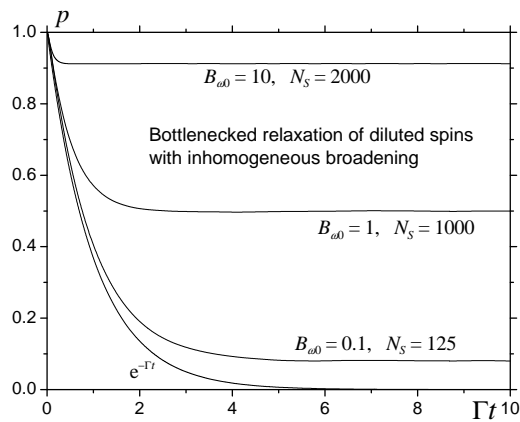}
\caption{Time dependence of the spin excitation $p(t)$ for the spins with
frequencies around $\protect\omega _{0}$ in the case of strong inhomogeneous
broadening.}
\label{Fig-Relaxation-inhomo}
\end{figure}

\begin{figure}[t]
\includegraphics[angle=-90,width=8cm]{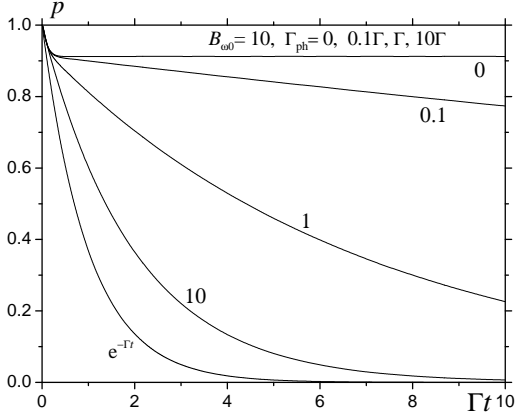}
\caption{The same in the case of damped phonons. Again, for $B\gg 1$ the
effective spin relaxation rate is much smaller than the phonon relaxation
rate $\Gamma _{\mathrm{ph}}.$}
\label{Fig-Relaxation-inhomo-damped}
\end{figure}

Next we calculated the time dependence of the spin excitation $p(t)$ for
spins within a frequency interval $\Delta \omega _{0}$ around $\omega _{0}$
that satisfies $\max (\Gamma ,\Delta )\ll $ $\Delta \omega _{0}\ll \delta
\omega _{0},$ as explained above. Again, the incoherent initial condition
for spins leading to Eq.\ (\ref{pinfRandomPhase}) was used. Only spins and
phonon modes within the interval $\Delta \omega _{0}$ were taken into
account while all other spins and phonon modes have been ignored. This
allowed to greatly reduce the computation time. The results have been shown
to be practically independent of $\Delta \omega _{0}$ as soon as the
condition $\Gamma \ll $ $\Delta \omega _{0}$ is fullfilled. The distribution
of spin frequencies within $\Delta \omega _{0}$ was taken equidistant
(similarly to the phonon modes) that allowed to eleminate statistical
scattering. In this realization of the model, spins and phonon modes form
two equivalent groups interacting with each other. The results of
computations for undamped phonons are shown in Fig.\ \ref
{Fig-Relaxation-inhomo}. Note that here oscillations visible in Fig.\ \ref
{Fig-Relaxation} are completely washed out. The asymptotic values of $p$ are
in accord with Eq.\ (\ref{pinfAvrRes}). Having the results for $p(t)$ for any
$\omega _{0},$ one could perform now integration over spin frequencies $%
\omega _{0}$ similarly to Eq.\ (\ref{pinfAvDef}) using, e.g., Eq.\ (\ref
{rhoSGaussian}).

One can compare the results of the present calculation within the frequency
interval $\Delta \omega _{0}$ with the results of the earlier general
calculation shown in Fig.\ \ref{Fig-Bottleneck_plateau-inhomo}. The
connection is provided by the identity
\begin{equation}
B_{\overline{\omega _{0}}}=\sqrt{\frac{\pi }{2}}\frac{\Gamma }{\delta \omega
_{0}}B  \label{BomegaviaB}
\end{equation}
that follows from Eqs.\ (\ref{BDef}), (\ref{Bomega0Def}), and (\ref
{rhoSGaussian}). For instance, the rightmost point of the curve $B=1$ in
Fig.\ \ref{Fig-Bottleneck_plateau-inhomo} is $p_{\infty }\simeq 0.070$ for $%
\delta \omega _{0}/\Gamma =30.$ Eq.\ (\ref{BomegaviaB}) yields then $B_{%
\overline{\omega _{0}}}\simeq 0.0418.$ For such small $B_{\overline{\omega
_{0}}}$ one can use Eq.\ (\ref{pinfomegaBomegasmall}) that yields $p_{\infty
}\simeq 0.030.$ The disagreement with the value $p_{\infty }\simeq 0.070$
can be explained by the fact that the number of spins $N_{S}=64$ in the
general calculation for $B=1$ is too small to reach the asymptotic result
given by Eq.\ (\ref{pinfAvrRes}) and statistical scattering is still
substantial. For the rightmost point of the curve $B=3$ in Fig.\ \ref
{Fig-Bottleneck_plateau-inhomo} the disagreement between the results of the
two calculations is smaller.

The results for spin relaxation in the case of damped phonons are shown in
Fig.\ \ref{Fig-Relaxation-inhomo-damped}. One can see that for $B\gg 1$ the
effective spin relaxation rate is much smaller than the phonon relaxation
rate $\Gamma _{\mathrm{ph}},$ similarly to the case without inhomogeneous
broadening, see Fig.\ \ref{Fig-Relaxation-damped}.

\section{Implementation for molecular magnets}

\label{Sec-parameters}

Let us work out the general expressions and estimate the parameters that
govern the bottlenecked spin relaxation for molecular magnets, in
particular, for the most popular compound Mn$_{12}.$ Historically, the
phonon botteneck was first observed in other systems. However, there is an
experimental evidence of the phonon bottleneck in molecular magnets as well.
On the other hand, molecular magnets are especially convenient because of
the universal form of the spin-phonon relaxation that does not depend on any
unknown spin-phonon coupling constants.\cite{garchu97prb} The spin
relaxation between the adjacent levels of the uniaxial spin Hamiltonian $%
-DS_{z}^{2}$ is due to the rotation of the crystallographic easy axis by the
transverse phonons. The rate of decay from the first excited state $\left|
-S+1\right\rangle $ to the ground state $\left| -S\right\rangle $ is given
by the  formula
\begin{equation}
\Gamma =\frac{S(2S-1)^{2}D^{2}\omega _{0}^{3}}{12\pi \hbar \rho v_{t}^{5}},
\label{GammaSSminus1}
\end{equation}
where $\rho $ is the mass density and $v_{t}$ is the speed of the transverse
phonons. The most recent derivation of this formula for the transitions
between any adjacent spin levels can be found in Appendix A of Ref.
\onlinecite
{chugarsch05prb}. In zero field one has $\hbar \omega
_{0}=E_{-S+1}-E_{-S}=(2S-1)D$ and Eq.\ (\ref{GammaSSminus1}) can be cast into
the elegant form
\begin{equation}
\Gamma =\frac{S}{12\pi }\frac{\omega _{0}^{5}}{\Omega _{t}^{4}},\qquad
\Omega _{t}\equiv \left( \frac{\rho v_{t}^{5}}{\hbar }\right) ^{1/4}.
\label{GammaSSminus1elegant}
\end{equation}
The phonon density of states of Eq.\ (\ref{rhoomegaDef}), multiplied by the
number 2 of transverse phonon modes, has the form
\begin{equation}
\rho _{\mathrm{ph}}(\omega _{0})=\frac{1}{\pi ^{2}}\frac{\omega _{0}^{2}}{%
\widetilde{\Omega }_{D}^{3}},\qquad \widetilde{\Omega }_{D}\equiv \frac{v_{t}%
}{v_{0}^{1/3}},  \label{rhophworkedout}
\end{equation}
where $v_{0}$ is the unit-cell volume. The frequency $\widetilde{\Omega }_{D}
$ is related to the Debye frequency $\Omega _{D}$ as $\Omega _{D}=(6\pi
^{2})^{1/3}\widetilde{\Omega }_{D}.$ Thus the bottleneck parameter $B$ of
Eq.\ (\ref{BDef}) becomes
\begin{equation}
B=\frac{12\pi ^{2}}{S}\frac{\Omega _{t}^{4}\widetilde{\Omega }_{D}^{3}}{%
\omega _{0}^{7}}n_{S}.  \label{BforMn12}
\end{equation}
For Mn$_{12}$ one has $S=10,$ $\rho =1.83$ g/cm$^{3},$ $v_{0}=3716$ \AA $%
^{3},$ and from the heat-capacity measurements\cite{gometal98prb} follows $%
\hbar \Omega _{D}/k_{B}\simeq 38$ K thus $\hbar \widetilde{\Omega }%
_{D}/k_{B}\simeq 10$ K. Further one obtains $v_{t}\simeq 2\times 10^{3}$ m/s
and $\hbar \Omega _{t}/k_{B}\simeq 210$ K, whereas $\hbar \omega
_{0}/k_{B}\simeq 12$ K. Plugging these parameters into Eq.\ (\ref{BforMn12})
one obtains $B\simeq 5\times 10^{5}n_{S}.$ This means that for a non-diluted
Mn$_{12}$ crystal, $n_{S}=1,$ the bottleneck parameter is huge.

Of course, for non-diluted magnetic crystals the physics includes the
effects of coherence and it is more complicated than just the phonon
bottleneck. The pure bottleneck situation is realized for a sufficient
dilution, so that $k_{0}r_{0}\gtrsim 2\pi $ \ and the phases of emitted and
reabsorbed phonons can be considered as random. Using $%
r_{0}=(v_{0}/n_{S})^{1/3}$ for the average distance between the neighboring
magnetic molecules and $k_{0}=\omega _{0}/v_{t},$ one can rewrite the
condition of sufficient dilution as
\begin{equation}
n_{S}\lesssim n_{S}^{\ast }\equiv \left( \frac{\omega _{0}}{2\pi \widetilde{%
\Omega }_{D}}\right) ^{3}.  \label{nStildeDef}
\end{equation}
We call $n_{S}^{\ast }$ critical concentration or critical dilution. With
the parameters above, one obtains $n_{S}^{\ast }\simeq 0.008.$ Even at this
dilution, the bottleneck parameter remains huge,
\begin{equation}
B^{\ast }\equiv \frac{12\pi ^{2}}{S}\frac{\Omega _{t}^{4}\widetilde{\Omega }%
_{D}^{3}}{\omega _{0}^{7}}n_{S}^{\ast }=\frac{3}{2\pi S}\left( \frac{\Omega
_{t}}{\omega _{0}}\right) ^{4}  \label{Bstar}
\end{equation}
that numerically yields $B^{\ast }\simeq 4000.$ Note that for spin
transitions between excited levels the energy differences $\hbar \omega
_{0}=E_{m+1}-E_{m}=-\left( 2m+1\right) D$ are smaller than above, thus the
values of $B$ are even larger.

Now from Eq.\ (\ref{Gammaviarho}) one obtains the estimation of the
spin-phonon matrix element
\begin{equation}
\left| V\right| ^{2}=\frac{\Gamma }{2\pi \rho _{\mathrm{ph}}(\omega _{0})}=%
\frac{S}{24}\frac{\omega _{0}^{3}\widetilde{\Omega }_{D}^{3}}{\Omega _{t}^{4}%
}.  \label{V2ResMn12}
\end{equation}
(that also could be obtained directly!). This yields the spin-phonon
splitting of Eq.\ (\ref{Deltansk}) in the form
\begin{equation}
\Delta =\sqrt{n_{S}}\sqrt{\frac{S}{6}\frac{\omega _{0}^{3}\widetilde{\Omega }%
_{D}^{3}}{\Omega _{t}^{4}}}  \label{DeltaMn12}
\end{equation}
and
\begin{equation}
\Delta ^{\ast }=\sqrt{n_{S}^{\ast }}\sqrt{\frac{S}{6}\frac{\omega _{0}^{3}%
\widetilde{\Omega }_{D}^{3}}{\Omega _{t}^{4}}}=\sqrt{\frac{S}{6}}\frac{1}{%
\left( 2\pi \right) ^{3/2}}\frac{\omega _{0}^{3}}{\Omega _{t}^{2}}.
\label{DeltaStarMn12}
\end{equation}
Numerically one obtains $\hbar \Delta /k_{B}\simeq 40\sqrt{n_{S}}$ mK and $%
\hbar \Delta ^{\ast }/k_{B}\simeq 3.6$ mK.

Let us now discuss the role of inhomogeneous broadening on the phonon
bottleneck in molecular magnets. First, there is the dipole-dipole
interaction (DDI) that is as strong as about $E_{DDI}/k_{B}\simeq 67$ mK
between the two neighboring Mn$_{12}$ molecules. This would result in the
change of the spin transition frequency for adjacent spin levels by $\left(
E_{DDI}/k_{B}\right) /S\simeq 6.7$ mK. Sometimes, for simplicity, the DDI is
considered as a kind of inhomogeneous broadening. This is an
oversimplification, at least for weakly excited states considered here.
Rigorous treatment of the Landau-Zener effect at fast sweep with an account
of both DDI and true inhomogeneous broadening (random hyperfine fields)\cite
{garsch05prb} shows that these two effects compete with each other, rather
than simply add. In the present case, the magnetostatic field smoothly
varies along the crystal (if the crystal shape is non-elliptic), so that its
change on the unit-cell distance is very small. Thus the DDI as a source of
inhomogeneous broadening will be neglected.

The greatest source of inhomogeneous broadening in Mn$_{12}$ is the
hyperfine interaction with their own $N_{I}=12$ nuclear spins $I=5/2.$ With
the hyperfine coupling $A$ between the total electronic spin $S$ and each of
the nuclear spins $I$ of $A/k_{B}=2$ mK (Ref. \onlinecite{harpolvil96}), the
dispersion of the hyperfine field $\delta H_{HF}$ on the electronic spin is
given by\cite{garchu97prb,garchusch00prb}
\begin{equation}
\delta H_{HF}=\frac{\sqrt{\sigma _{I}}A}{g\mu _{B}},\qquad \sigma _{I}=\frac{%
N_{I}}{3}I(I+1),  \label{deltaHF}
\end{equation}
whereas the inhomogeneous line shape is Gaussian. Numerically one obtains $%
\delta H_{HF}\simeq 8.8$ mT. For the transition between adjacent spin levels
for the inhomogeneous broadening $\delta \omega _{0}$ in Eq.\ (\ref
{rhoSGaussian}) one has $\hbar \delta \omega _{0}=g\mu _{B}\delta H_{HF},$
thus $\hbar \delta \omega _{0}/k_{B}\simeq 12$ mK. In another popular
compound Fe$_{8}$ the inhomogeneous broadening is mainly due to the DDI with
nuclear spins of hydrigen atoms present in magnetic molecules, $\delta
H\simeq 0.8$ mT according to Ref. \onlinecite{weretal00prl}.

One can see that in Mn$_{12}$ the inhomogeneous broadening 12 mK is smaller
than the spin-phonon gap $\hbar \Delta /k_{B}\simeq 40$ mK in the
non-diluted case, thus it can be neglected in the first approximation. On
the other hand, as said above, the pure bottleneck case requires dilution
below $n_{S}^{\ast }$ of Eq.\ (\ref{nStildeDef}), and for $n_{S}\lesssim $ $%
n_{S}^{\ast }$ one has $\delta \omega _{0}>\Delta .$ For Fe$_{8}$ the
inhomogeneous broadening begins to play a role for a very strong dilution
where measurements are difficult. If the inhomogeneous broadening is
dominating, the relaxation is governed by the frequency-interval botleneck
parameter $B_{\overline{\omega _{0}}}$ of Eq.\ (\ref{BomegaAvrGauss}). \ With
the help of Eq.\ (\ref{rhophworkedout}) one obtains
\begin{equation}
B_{\overline{\omega _{0}}}=\frac{n_{S}}{\sqrt{2\pi }}\frac{\pi ^{2}%
\widetilde{\Omega }_{D}^{3}}{\overline{\omega _{0}}\delta \omega _{0}}.
\label{BomegaAvrGauss1}
\end{equation}
At the critical dilution this becomes
\begin{equation}
B_{\overline{\omega _{0}}}^{\ast }=\frac{n_{S}^{\ast }}{\sqrt{2\pi }}\frac{%
\pi ^{2}\widetilde{\Omega }_{D}^{3}}{\overline{\omega _{0}}\delta \omega _{0}%
}=\frac{1}{2^{7/2}\pi ^{3/2}}\frac{\overline{\omega _{0}}}{\delta \omega _{0}%
}\cong 0.016\frac{\overline{\omega _{0}}}{\delta \omega _{0}}.
\label{BstaromegaAvr}
\end{equation}
That is, for any line with a well-defined central frequency $\overline{%
\omega _{0}}$ the bottleneck parameter is large. With the above parameters
for Mn$_{12}$ one obtains $B_{\overline{\omega _{0}}}\simeq 2\times 10^{3}$ $%
n_{S},$ $\overline{\omega _{0}}/\delta \omega _{0}\simeq 1000,$ and $B_{%
\overline{\omega _{0}}}^{\ast }\simeq 16$ that is still large. Then Eq.\ (\ref
{pinfInhomoLargeB}) yields $1-p_{\infty }\simeq 0.11,$ that is, only a 10\%
of the initial spin excitation can relax before the bottleneck plateau is
reached. Thus the inhomogeneous broadening in Mn$_{12}$ does not fully
resolve the PB, and in Fe$_{8}$ its effect is even smaller.

\section{Discussion}

\label{Sec-Discussion}

The problem of the phonon bottleneck considered in this paper in the
weak-excitation limit is only a part of a larger problem of collective
spin-phonon relaxation. The bottleneck in the pure form occurs for
magnetically diluted systems that satisfy the condition $k_{0}r_{0}\gg 1,$
where $k_{0}$ is the wave vector of a resonant phonon and $r_{0}$ is the
typical distance between the neighboring spins. If this condition is
violated, one cannot consider the phases of emitted phonons reaching other
spins as random, and the interference effects become important. The
well-known example of interference effects is superradiance \cite
{dic54,chugar04prl} that requires, among other conditions, a coherent
initial condition of spins. In contrast to the bottleneck, superradiance
dramatically increases the relaxation rate, so that the two effects should
compete. On the other hand, destructive interference effects in the case of
incoherent initial condition or dynamical loss of coherence due to the
inhomogeneous broadening can lead to suppression of relaxation that
resembles the PB but has a different physical origin. For this reason, using
the results of this paper to interpret experiments should be done with care
as the experimental observations can be a mixture of different effects. In
particular, the results obtained in this paper are not applicable to
non-diluted molecular magnets.

Considering the weak-excitation limit in this paper allowed to drastically
simplify the Schr\"{o}dinger equation and obtain numerically exact results
for the PB effect with and without inhomogeneous broadening of spins and
phonon damping. It was confirmed that the bottleneck parameter $B$
quantifying the statistical weights of spins and resonant phonons,
introduced in Ref. \onlinecite{gar07prb}, plays the main role in the
problem. Fundamentally the most interesting case corresponds to the pure
model without inhomogeneous spin broadening and phonon damping. For this
model, similarly to the results of Ref. \onlinecite{gar07prb}, the spin
excitation $p(t)$ was shown to oscillate approaching the bottleneck plateau,
see Fig.\ \ref{Fig-Relaxation}. However, these oscillations have a smaller
amplitude and are stronger damped than the analytical results of Ref. %
\onlinecite{gar07prb}. The frequency of these oscillations corresponds to
the gap $\Delta $ between the two branches of the hybridyzed magnetoelastic
waves at resonance [see Eq.\ (\ref{OmegapmDef}) and Fig.\ \ref{Fig-splitting}]

Here it was shown that the bottleneck parameter effectively decreases in the
presence of inhomogeneous broadening that alleviates the bottleneck
condition. If the inhomogeneus spin line width $\delta \omega _{0}$ exceeds
the spin-phonon gap frequency $\Delta /\hbar ,$ the splitting of spin and
phonon modes is not resolved and the oscillations of $p(t)$ are washed out,
see Fig.\ \ref{Fig-Relaxation-inhomo}.

Inclusion of the \emph{ad hoc} phonon relaxation rate $\Gamma _{\mathrm{ph}}$
in the theory describes the second, post-plateau, stage of the spin
relaxation. An important observation is that for $B\gg 1$ the corresponding
relaxation rate is much smaller that $\Gamma _{\mathrm{ph}}$ (see Figs.\ \ref
{Fig-Relaxation-damped} and \ref{Fig-Relaxation-inhomo-damped}) and the spin
relaxation is non-exponential [see Eq.\ (\ref{ptSpreadRandom2DampedSecond})
and Fig.\ \ref{Fig-Relaxation-damped-log}]. One should stress, however, that
it is not completely satisfactory to plug an \emph{ad hoc} phonon relaxation
into the theory. The latter should be treated in this case as taken from the
experiment. This can lead to a problem since one of the main sources of the
observed phonon damping can be their scattering on spins that is already
taken into account by the very spin phonon interaction, Eq.\ (\ref{VRWA}). On
the other hand, this and other kinds of elastic scattering cannot help the
system to reach complete equilibrium since these processes conserve the
energy and do not transfer excitation from the narrow group of resonant
phonons to the rest of the phonon bath.

It would be of a principal importance to generalize the theory of phonon
bottleneck for the higly excited initial states of the spin subsystem.
However, the Schr\"{o}dinger equation in this case is not amenable to a
direct numerical solution. In Ref. \onlinecite{chugar04prl} it was argued
that, since collective motion of spins in the regions large in comparizon to
the typical distance between the neighboring spins involves a large number
of atomic spins, the problem can be considered classically. Thus the spin
and phonon operators in Eqs.\ (9) and (10) of Ref. \onlinecite{chugar04prl}
had been replaced by classical variables. This gives a possibility to
numerically treat highly excited states of the spin-phonon system without
the combinatorial explosion of the full SE. On the other hand, linearization
of the classical equations of motion of Ref. \onlinecite{chugar04prl} near
the ground state yields equations that are equivalent to the truncated SE,
Eq.\ (\ref{SELowExc}), in the low-excitation limit. This suggests that
classical equations Ref. \onlinecite{chugar04prl} are indeed a good
approximation for the spin-phonon problem in the whole energy range. It
would be interesting to investigate what are the quantum corrections to
these equations.

\begin{acknowledgements}
This research is supported by the PSC-CUNY grant
 PSCREG-38-276. Numerous stimulating discussions with
E. M. Chudnovsky and A. Kent are greatfully acknowledged. \end{acknowledgements}

\bibliography{gar-own,gar-oldworks,gar-relaxation,gar-tunneling,gar-books}

\end{document}